\documentclass[preprint,prd,aps]{revtex4}
\usepackage{axodraw}
\begin{document}
\newcommand{\DVslash}{\line(4,1){22} \!\!\!\!\!\!\!\!\!\!\Delta V}


\title{Renormalization in Reparameterization Invariance}

\author{Yu-Qi Chen}
\affiliation{Institute of Theoretical Physics, Academia Sinica,
Beijing 100080, P.R. of China }

\begin{abstract}
The renormalization issue of the reparameterization invariance in
heavy quark effective theory and NRQCD is investigated. I argue
that the renormalization of the transformation of the heavy quark
field under the variation of the velocity parameter $V$ is
attributed to the renormalization of the small component field in
the proposed transformation. I show that the matching condition
for determining the renormalized small component field can be
obtained by imposing an infinitesimal transformation of $V$ on the
relations between the Green's functions in QCD full theory and
those in the effective theory. As an application, I determine the
renormalized transformation to order $1/m^2$ using the matching
condition. The obtained result is in disagreement with that
determined by indirect method.
\end{abstract}

 \vfill \eject
\maketitle

\section{introduction}
Heavy quark effective theory (HQET)\cite{HQET} and nonrelativistic
QCD (NRQCD)\cite{NRQCD} are powerful tools in dealing with
dynamics of heavy-light and heavy-heavy systems, respectively. In
those systems, the off-shell momentum of the heavy quark is much
smaller than its mass. The effective theories are designed to
reproduce the results of the QCD full theory at the low energy
scale in a simpler way by integrating out the effects at the
energy scale of the heavy quark mass.  In the past decade both
effective theories and their applications have been intensively
studied.

One interesting theoretical issue in those effective theories is
the reparameterization invariance (RPI). It arises from the fact
that the effective theory explicitly depends on the four velocity
parameter $V$.  In constructing the effective lagrangian, one
needs to divide the heavy quark momentum $P$ into a large part and
a small one as $P=mV +k$, where $m$ is the heavy quark mass and
$k$ is a small residual momentum. one also needs to decompose the
Dirac 4-fermion field as large and small two component fields in
respect of $V$ and uses the large one to describe the heavy quark
or antiquark. These procedures lead to the effective lagrangian
being $V$-dependent. The choice of $V$ which satisfies $V^2=1$ is
not unique. But the physical prediction should be unchanged
against the variation of the velocity parameter $V$. This is the
RPI. It is required by the consistence of the effective theory and
also conducts interesting applications. It was first proposed in
HQET. However, the same invariance also holds in NRQCD effective
theory.

To implement RPI in the effective theory, it is essential to find
out an appropriate transformation of the heavy quark field under
the variation of $V$,  which is found to be quite nontrivial. It
was first studied by Manohar and Luke\cite{Luke:1992cs} in HQET.
They used the Lorentz boost of spinor field as the transformation
of the heavy quark field from finite velocity $V\to V'$. Their
transformation suffers from operator ordering ambiguities when it
is expanded to a higher order of $1/m$ while
Manohar\cite{Manohar:1997qy} discussed its expansion at higher
order. Chen\cite{Chen:sx} proposed an infinitesimal transformation
of the heavy quark field under the velocity variation  from $V \to
V + \Delta V$. Chen's transformation keeps tree level effective
lagrangian invariant to all orders of $1/m$. Finkemeier, Georgi,
and McIrvin\cite{Finkemeier:1997re} showed that to order $1/m^2$
the effective lagrangian constrainted  by Manohar and Luke's
transformation and Chen's transformation may be related by a field
redefinition.

 Chen's transformation can be expanded as inverse power series of
the heavy quark mass. Each term contains  the product of  some
covariant derivatives and the heavy quark field. Since this kind
of expansion changes the ultraviolet behavior of the original
transformation, beyond the next leading order, it turns out that
it  needs to be renormalized. The renormalization of it is
different from that of the effective lagrangian in which each term
is the bilinear function of the heavy quark fields. For that case,
we have appropriate matching conditions to determine the
renormalized transformation.

 Kilian and Ohl\cite{Kilian:1994mg} proposed a renormalized
transformation. The form is exactly the same with  Chen's
transformation except that  the covariant derivative $D^\mu$ in
Chen's transformation is substituted by another operator which
they called as the general covariant derivative.
Sundrum\cite{Sundrum:1997ut} discussed this issue using the
auxiliary field method and obtained a result similar with Kilian
and Ohl's. The results presented in these papers are formal. They
did not showed how to determine the general covariant derivative
by some specific matching conditions. Actually, in literatures no
specific calculation for determining the transformation has been
done with this method. The only calculation to determine the
renormalized transformation is given by Balzereit in an
unpublished paper\cite{Balzereit}. He first calculated the
effective lagrangian to order $1/m^3$ at one loop level in the
leading logarithmic approximation. By requiring the effective
lagrangian invariant, he then determined the renormalized
transformation to order $1/m^2$ indirectly. There are some
drawbacks in this kind of calculations. First, they are quite
complicated  since the determination  of  higher order effective
lagrangian usually is hard work. Second, it makes the RPI less
practical application. An interesting application of the RPI is
that once we know the transformation, we can use it to constraint
the higher order effective lagrangian and makes the calculations
simpler\cite{Luke:1992cs,Chen:sx,Finkemeier:1997re}. Balzereit did
it in an inverse order so RPI may not be used to constraint the
effective lagrangian in his method.

In this paper, I study this issue in an alternative way, with
emphasizing on  determining the renormalized transformation using
the matching conditions. It is essential in studying the
renormalization issue of RPI. With it, one is able to determine
the renormalized transformation to any desired order both in
$\alpha_s$ and in $1/m$ expansion.

To derive the matching condition, it is instructive to recall the
renormalization procedure of the effective lagrangian. There are
simpler relations between the Green's function in the unexpanded
effective theory with nonlocal form. These relations ensure that
the effective theory and the full QCD reproduce the same physical
predictions. The expanded effective theory changes the ultraviolet
behavior of the unexpanded effective theory. In order to reproduce
the same result of the full QCD, those relations are required to
be satisfied and one has to add some local operators in the
effective lagrangian, i.e., renormalizing the effective
lagrangian. Therefore, those relations between the Green's
functions can be used as the matching conditions\cite{HQET} to
determine the effective lagrangian. On the other hand, there is no
direct relations between the 1PI vertexes in the effective theory
and those in the full QCD. Thus one may not obtain the matching
conditions from the 1PI vertex.

As we will see later on,  the renormalization of the
transformation of the heavy quark field under the variation of the
velocity parameter $V$ is attributed to the renormalization of the
small component field in the proposed transformation. Therefore,
if we can find some relations between certain correlation
functions in the full QCD and those in the effective theory, which
involve the small component field, it then can be used as the
matching conditions to determine the renormalized transformation.
It is found that imposing an infinitesimal transformation on both
sides of the relations between the Green's functions in  QCD full
theory and those in the effective theory, new relations in which
the small component field is involved on the effective theory side
are obtained.
 These new relations are nothing but the matching
conditions to determining the renormalized small component field.
They provide a systematic way to determine the renormalized
transformation.  As a specific example, I use these matching
conditions to determine the renormalized transformation to order
$1/m^2$ at one-loop level. The obtained result is in disagreement
with that obtained by Balzereit\cite{Balzereit}. Since the same
lagrangian (MRR lagrangian) are used, the disagreement between
these two different results cannot be accounted by a field
redefinition. I then show that the renormalized transformation
determined by these matching conditions can be written in the form
of Chen's transformation with the covariant derivative substituted
by an operator which may be called as a general covariant
derivative. Thus renormalized transformation determined by the
matching conditions presented in this paper is consistent with
Kilian and Ohl's result.\cite{Kilian:1994mg}

The remainder of the paper is organized as follows. In section II,
after a brief review on the tree level transformation, I argue
that the renormalized transformation of the heavy quark field is
attributed to the renormalization of the small component field. I
then show that the matching condition for determining it can be
obtained by imposing an infinitesimal transformation on the
general relations between the Green's functions in  QCD full
theory  and those in the effective theory. As an example, in
section III, I determine the renormalized transformation to order
$1/m^2$ by matching the two-point and the three-point functions.
In section IV, I show that previous results can be more clearly
understood in an alternative way to construct the effective
lagrangian, where a four-component effective lagrangian
constructed first, followed by its reduced into the effective
lagrangian in a two-component field. I then show that the
remormalized small component field determined by the matching
conditions are consistent with Kilian and Ohl's
result.\cite{Kilian:1994mg}. Then I shown that the remormalized
effective lagrangian is reparameterization invariant under the
renormalized transformation. Section V contributes to the
conclusion. Finally, in appendix A, I derive the general relation
between the Green's functions using the generating functional
method.


\section{renormalized transformation of the heavy quark field}

 In the heavy quark effective theory, the heavy quark is described
by a two component field while in QCD full theory it is described
by the Dirac four-component field. Thus, to construct the
effective theory, one needs first to decompose the Dirac
four-component field into two-component field. A simpler way to
realize this decomposition introduced by Georgi\cite{HQET} is:
\begin{equation}
 h_{V\pm}(x)\;\equiv\;\exp\left(\,imV\cdot x\,\right)\;
 \mathrm{P}_{\pm}\,\Psi(x)\;,
 \label{hvp}
\end{equation}
where
\begin{equation}
\mathrm{P}_{\pm} \;\equiv \; \frac{1\,\pm\, \not\!{V}}{2}\;,
 \label{ppm}
\end{equation}
are the projection operators. The phase factor just removes the
large part of the heavy quark momentum when it is written as $p=mV
+k$. This definition of the field is used by most
people\cite{HQET} in literatures. Nevertheless, it is not unique.
Different definitions lead to different forms of the effective
lagrangian. However, they can be related to each other by field
redefinition and produce the same physical
predictions\cite{Finkemeier:1997re}. Overall in this paper, we use
the definition of (\ref{hvp}).

\subsection{the tree level transformation}
With the definition (\ref{hvp}), the effective lagrangian
reads\cite{Mannel:1991mc,Chen:sx}:
%
\begin{eqnarray}
{\cal L}_{\rm eff}^0 \;&=&\;
 \bar h_{V+} (x)\,i D \cdot V \,h_{V+}(x)\,
 -\, \bar h_{V+}(x)\, \not\! {D}\,
 \frac{1}{2m+iD \cdot V}\, P_{V-} \not\!{D}\, h_{V+}(x)
 \;,
  \label{L-tree}
\end{eqnarray}
where $D^\mu\equiv \partial^\mu -i g_s A^\mu $ is the covariant
derivative. This is the nonlocal form of the effective theory.

Obviously, this effective lagrangian depends on the velocity
parameter $V$. The choice of this parameter is not unique. The RPI
implies that physical predictions by the effective theory are
independent of the choice of $V$. In Ref.\cite{Chen:sx},  it was
shown that the effective lagrangian (\ref{L-tree}) is invariant
under an infinitesimal transformation $V\rightarrow V+ \Delta V$
\begin{eqnarray}
\Delta h_{V+}(x) \;&=&\; \frac{\DVslash}{2} \,
  \Big(\,   h_{V+}(x) \,+\,  h_{V-}(x) \,\Big)
 \;.
  \label{D-V}
\end{eqnarray}
with the $h_{V-}(x)$ being the small component field and given by:
\begin{eqnarray}
 h_{V-}(x) \;&=&\; \frac{1}{2m+iD\cdot V}\, \mathrm{P}_{-} \,i \not\! D
 \,h_{V+}(x)\;.
 \label{hm-tree}
\end{eqnarray}
$\Delta V$ is constrained by $\Delta V\cdot V =0$ due to $V^2=1$.

Both the effective lagrangian (\ref{L-tree}) and the
transformation given by (\ref{D-V}) and (\ref{hm-tree})  can  be
expanded as power series of $1/m$. The RPI is then valid order by
order in $1/m$. It implies that the tree level transformation
makes the tree level effective lagrangian valid at any order of
$1/m$. The cancellation of  the lagrangian shift $\Delta L$ at
each order is quite nontrivial. Without expansion, the effective
theory with the lagrangian (\ref{L-tree}) is equivalent to that of
the full theory in the sense that they produce the same $S-$matrix
elements.

\subsection{matching condition for renormalizing the transformation}

With expansion in terms of  $1/m$, the ultraviolet behavior of the
theory is changed. To compensate this change, both the effective
lagrangian and the transformation receive renormalization. The
renormalization procedure of the effective lagrangian is
well-known while it is not evident how to renormalize the
transformation. The difficulty arises from the fact that each term
contains the product of some covariant derivatives and the heavy
quark field which is complicated and it is just the linear
function of the heavy quark field but not bi-linear functions as
those in the effective lagrangian.

Nevertheless, to gain the answer, it is instructive to recall the
renormalization procedure of the effective lagrangian. For the
heavy quark field defined in (\ref{hvp}), there are simpler
relations between Green's functions in the effective theory with
the nonlocal form (\ref{L-tree}) and those in QCD full theory. For
the expanded effective lagrangian, one has to add some local
operators to the expanded effective lagrangian so that those
relations can be satisfied. Consequently, the renormalized
effective lagrangian can reproduce the results of full QCD.
Therefore, those relations between the Green's functions are just
the matching conditions to determine the effective lagrangian.

In order to renormalize the expanded transformation, the key point
is to gain appropriate matching conditions to determine the
transformation. As we will see, the renormalization of the
transformation of the heavy quark field under the variation of the
velocity parameter $V$ is attributed to the renormalization of the
small component field in the proposed transformation (\ref{D-V}).
Therefore, if we can establish some relations between the Green's
functions in QCD full theory and certain correlation functions
which involve the $h_{V-}(x)$ field in the effective theory, they
can be used as the matching conditions to renormalize the
$h_{V-}(x)$ and hence the transformation. It is found that these
relations can be obtained by imposing an infinitesimal
transformation of $V$ on both sides of the relations between the
Green's functions in QCD full theory and those in the effective
theory.

Now let's first look at the relations between the Green's
functions in the full QCD and those in the effective theory.

Denote the Green's function in the full theory by $G(x,y;B)$ and
that in the effective theory by $G_V(x,y;B)$, respectively, where
$B$ is an arbitrary background field.  They are defined by
\begin{eqnarray}
 G(x,y;B) \; \equiv \; \langle \,0\, |\, T\,\Psi(x)\,\bar{\Psi}(y)\,|\, 0 \,
 \rangle^B\;,
 \label{G-f}
\end{eqnarray}
and
\begin{eqnarray}
 G_V(x,y;B) \;\equiv \;\langle \, 0\, |\,T \,h_{V+}(x)
 \,\bar{h}_{V+}(y)\,|\, 0 \,\rangle^B\;,
 \label{G-e}
\end{eqnarray}
Any interaction vertex with gluon can then be obtained by
functional differentiating over $B(x)$. When the quark field
$h_{V+}(x)$ is related to the field in the full theory by
Eq.(\ref{hvp}), the Green's functions satisfy the following simple
relation (A derivation of this relation using generating
functional method is given in Appendix A):
\begin{eqnarray}
 G_V(x,y;B) \;& \doteq &\; \mathrm{P}_{+} \, G(x,y;B) \, \mathrm{P}_{+}\;,
 \label{G-Gv}
\end{eqnarray}
here $\doteq$ means that we omit the phase factor $\exp{[i m
V\cdot (x-y)]}$ and the renormalization constant
$Z(m,\alpha_s(m))$ which arises from the renormalization of the
heavy quark field in the full theory and that in the effective
theory. Both sides are valid to all orders of $1/m$ and
$\alpha_s$. This relation ensures that the $S-$matrix elements in
the effective theory are identical to those in the full theory.
The local effective theory is gained by expanding the nonlocal
effective lagrangian density (\ref{L-tree}). This expansion
changes the ultraviolet behavior of the nonlocal effective theory.
To reproduce the  result of the nonlocal effective theory, one
needs to add local operators to the expanded effective lagrangian.
The relation (\ref{G-Gv}) is required to be satisfied as matching
conditions to determine the coefficients of those local operators.
Thus the relations between the Green's functions in the full
theory and those in the effective theory can be used as the
matching conditions to determine the effective lagrangian.

Below starting from these relations we derive the matching
conditions to determine the renormalized transformation of the
heavy quark field.

 Obviously, these relations are valid for arbitrary $V$. It allows us impose
an infinitesimal transformation $V\rightarrow V+ \Delta V$ on both
sides. It follows that:
\begin{eqnarray}
 \Delta G_V(x,y;B) \;& \doteq &\; \frac{\DVslash}{2} \, G(x,y;B) \, \mathrm{P}_{+} +
  \mathrm{P}_{+} \, G(x,y;B) \, \frac{ \DVslash}{2}\;.
 \label{DG-DGv}
\end{eqnarray}
Again the symbol $\doteq$ means we omit the  phase factor and a
term arises from its infinitesimal shift which is trivial under
the transformation.

Given the definitions of the Green's functions in
Eqs.(\ref{G-f}),(\ref{G-e}), we have the following  unique
solution of (\ref{DG-DGv}):
\begin{eqnarray}
 \langle \,0\, |\,T\,\big(\,\Delta h_{V+}(x)\,\bar{h}_{V+}(y)\,\big)
 \,|\, 0\,\rangle^B
  \;&\doteq &\;
 \frac{\DVslash}{2} \,\langle \,0\, | \,T\,\Psi(x)\,\bar{\Psi}(y)\,|\,
 0\, \rangle^B
 \mathrm{P}_{+}\;.
 \label{D-h}
\end{eqnarray}

Eq.~(\ref{D-h}) implies that $\Delta h_{V+}(x)$ is proportional to
$\DVslash$. Thus we can generally write it as the following form
\begin{eqnarray}
 \Delta h_{V+}(x) \;&=&\; \frac{\DVslash}{2} \,
 \Big(\, \mathrm{P}_{+}\, h'_{V+}(x) \,+\, \mathrm{P}_{-}\,
 h'_{V-}(x)\, \Big) \;.
 \label{D-h-pm}
\end{eqnarray}
Substituting it into (\ref{D-h}), Eq.~(\ref{D-h}) is then
decomposed into two equations by projection operator:
\begin{eqnarray}
 \langle\, 0\, |\,T\,\big(\, h'_{V+}(x)\,\bar{h}_{V+}(y)\,\big)\,|\, 0\, \rangle^B
  \;&\doteq &\;
 \mathrm{P}_{+} \,\langle\, 0\, |\, T\,\Psi(x)\,\bar{\Psi}(y)\,|\, 0\,
 \rangle^B \,\mathrm{P}_{+}\,
 \label{hpp} \,, \\
 \langle \,0\, |\,T\,\big(\, h'_{V-}(x)\,\bar{h}_{V+}(y)\,\big)\,|\, 0\, \rangle^B
  \;&\doteq &\;
 \mathrm{P}_{-} \,\langle\, 0\, | \,T\,\Psi(x)\,\bar{\Psi}(y)\,|\, 0\, \rangle^B\,
 \mathrm{P}_{+}
 \label{hmp} \;.
\end{eqnarray}
Eq.~(\ref{hpp}) is nothing but Eq.~(\ref{G-Gv}) if $h'_{V+}(x)$ is
identical to $h_{V+}(x)$. Eq.~(\ref{hmp}) is a new one which can
be regarded as the definition of the $h'_{V-}(x)$. Some comments
can be given about this equation. The right hand side is still the
Green's functions in the full theory while the left hand side is
the Green's functions in the effective theory with insertion of
local operators at point $x$. Again both sides are valid at any
desired order of $1/m$ and  $\alpha_s$ with all possible
interaction with gluons.

For the effective lagrangian at tree level, this equation is
satisfied if $\Delta h_{V+}(x)$ is given by transformation
(\ref{D-V}) with $h_{V-}(x)$ given by the expanded expression
(\ref{hm-tree}). At loop level, it is also valid at any specific
loop momentum which is smaller than the quark mass. However, in
calculating the whole loop momentum integration, some differences
arise and the (\ref{hmp}) can be regarded as the matching
condition to determine the renormalized $h_{V-}(x)$ field.

Eq.~(\ref{D-h-pm}) together with  (\ref{hpp}) and (\ref{hmp})
imply that the renormalized transformation keeps the same form as
the tree-level transformation (\ref{D-V}). But only the small
component field needs to be renormalized.

Now let's illustrate how (\ref{hmp}) determine the renormalized
$h_{V-}(x)$ field as the matching conditions. For simplicity, we
limit our arguments in the hard cutoff regularization. Similar
arguments are applicable to dimensional regularization.

Suppose one takes different hard cut-off regularization energy
scales $\Lambda_e$ and $\Lambda_f$ in effective theory and in full
theory, respectively. The $\Lambda_f$ should be much larger than
the heavy quark mass $m$ for including both quark and antiquark
contributions. The $\Lambda_e$ should be much smaller than $m$ for
the validation of the $1/m$ expansion. So they satisfy a hierarchy
relation $\Lambda_e \ll m \ll \Lambda_f$. Thus in calculating loop
diagrams the integration bounds are different on both sides ( the
left hand side is integrated out from 0 to $\Lambda_e$ while the
right hand side is integrated from 0 to $\Lambda_f$ ). It leads to
that the tree level expression of $h_{V-}(x)$ given by the
expansion of (\ref{hm-tree}) makes relation (\ref{hmp}) no longer
valid. To compensate the differences, one has to add the
contributions of the loop momentum integrals from $\Lambda_e$ to
$\Lambda_f$ to the left hand side. Those contributions can be
expressed as the insertions of the local operators on the heavy
quark line. They may be at point $x$ or not. Those local operators
at point $x$ can be absorbed into the redefinition of the $
h_{V-}'(x) $ while those local operators not shrunk at the point
$x$ correspond to the insertions of the higher dimensional
operators in the effective lagrangian. Therefore,
Eq.~(\ref{D-h-pm}) is just the matching conditions for determining
the renormalized small component field $h_{V-}(x)$.
Eq.~(\ref{hmp}) allows one determine the renormalized $h_{V-}(x)$
to any order of $1/m$ and $\alpha_s$.

\section{ renormalized transformation to order $1/m^2$ }
As a specific example, in this section, we determine the
renormalized effective lagrangian and the $h_{V-}(x)$ field up to
the next leading order corrections  of $1/m$  using the matching
conditions (\ref{G-Gv}) and (\ref{hmp}), respectively. Here we use
the dimensional regularization and Feynman gauge in all specific
calculations.

Up to next leading order corrections of $1/m$, the most general
form of the renormalized effective lagrangian can be expressed as:
\begin{eqnarray}
L_{V}^1(x) \;&=&\; Z\,\bar h_{V+} (x)\,i D \cdot V\, h_{V+}(x)\,
  +\,{Z \over 2m} \,\bar h_{V+} (x)\, \mathbf{D}^2 \,h_{V+} (x)
  \nonumber\\ &&
  \;+\,{ZZ_e \over 2m} \,\bar h_{V+} (x)\, ({iD\cdot V})^2\, h_{V+} (x)
  \,+\,{ZZ_m \over 4m}\, h_{V+} (x)\, \sigma^{\mu\nu} G_{\mu\nu}(x)\, h_{V+} (x)
  \;,
\label{LV-1}
\end{eqnarray}
$\sigma^{\mu\nu} \equiv {i\over 2} [\gamma^\mu,\gamma^\nu]$,
 and the most general form of the renormalized
 $h_{V-}(x)$ field can be written as:
\begin{eqnarray}
 h_{V-}(x) \;&=&\; \mathrm{P}_{-}\, \Big( \,{d_0(\mu) \over 2m}
 \,i\not \!D \,+\, {d_1(\mu)\over 4m^2 }\,
   D\cdot V \not \!D \, +\, {d_2(\mu) \over 4m^2} \,\not \!D  D\cdot V \,\Big)
    h_{V+}(x)\;,
 \label{hm-1}
\end{eqnarray}
where $Z, Z_e, Z_m, d_0(\mu), d_1(\mu), d_2(\mu)$ are the short
distance coefficients to be determined.

In the following matching procedures, all the short-distance
coefficients are assumed to be calculated to all orders in
$\alpha_s$, which makes the matching procedure are applicable for
higher order calculations. But in this paper we only present
one-loop result in the final expression.

Since the off-shell momenta of the heavy quark and the momenta of
the gluons can be treated as much smaller than the quark mass, in
the matching procedure, the integrand can be expanded as power
series of these momenta over the heavy quark mass. It leads to
that the remainder part of the loop momentum integrals no longer
depends on the heavy quark mass. We are free to choose the
infrared regulator since the infrared divergences cancel on both
sides. If we use a limitation order in which the external
off-shell momenta of heavy quark and the momenta of the gluons go
to zero first, followed by $\epsilon=2-D/2$ going to zero, as used
by Eichten and Hill in Ref.\cite{HQET}, then all terms such as
$k^\epsilon$ arising from the loop momentum integrals vanish. In
the effective theory, this implies that all contributions from the
loop diagrams vanishes since all loop momentum integrals are
proportional to it while in the full theory it implies that there
is no any logarithmic nonlocal terms of these external momenta.
This simplifies the matching calculations significantly.

It is easy to see that to determine these coefficients, we need to
match both the two-point and three-point functions.

\subsection{matching two-point function}
Let's first see what we can learn from  matching the two-point
function.  For external momentum $p$ of the heavy quark near
mass-shell, the general form of the QCD heavy quark self-energy,
the inverse of the two-point function, with above infrared
regulator can be written as:
\begin{eqnarray}
\Sigma(p) \;&=&\; A\,(\not\!p-m) \;+\;{2B m \Delta}\;,
\label{sigma}
\end{eqnarray}
where $\Delta$ is defined as
\begin{eqnarray}\label{delta}
\Delta \;&\equiv&\; {p^2-m^2 \over 4m^2} \;.
\end{eqnarray}
The heavy quark expansion implies that $\Delta \ll 1$. Thus  $A$,
$B$ can be expanded as power series of $\Delta$. Up to next
leading order, they can be written as:
\begin{eqnarray}
A \;&=&\; 1\,+\,c_0(\mu) \,+\,c_2(\mu) \,\Delta \;,\nonumber \\
B \;&=&\; c_1(\mu) \,+\, c_4(\mu)  \,\Delta \;.
  \label{AB}
\end{eqnarray}

At one-loop level in QCD, there is  only one 1PI diagram
contributing to the self-energy, as shown in Fig. 1. Carrying out
a specific calculation with the above infrared regulator, we
obtain that:
\begin{eqnarray}
&& c_0(\mu)= {C_F\over 4\pi } \alpha_s(\mu)\Big( \ln{\mu^2\over
m^2}+2\Big )\,,
 ~~~~~~~~
 c_1(\mu)= {C_F\over 2 \pi} \alpha_s(\mu)\Big( \ln{\mu^2\over m^2}+1\Big )\,,
 \nonumber \\ &&
 c_2(\mu)=-{C_F \over \pi} \alpha_s(\mu) \Big( \ln{\mu^2\over m^2}+2\Big )\,,
 ~~~~~~
 c_4(\mu)= -{C_F\over  \pi} \alpha_s(\mu)\Big( \ln{\mu^2\over m^2}-1\Big
 )\,,
 \label{c0-4}
\end{eqnarray}
with $C_F=4/3$.
\bigskip\bigskip

    \begin{center}\begin{picture}(300,60)(0,0)
    \SetWidth{0.7}
    \SetColor{YellowOrange}
    \ArrowLine(80,20)(220,20)
    \SetColor{Cerulean}
    \GlueArc(150,20)(40,0,180){5}{8}   
   \SetColor{Brown}
      \Vertex(190,20){2} \Vertex(110,20){2}
   \end{picture}\\
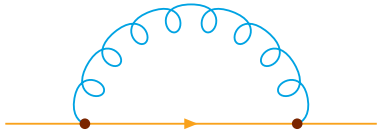
    {\footnotesize Fig. 1 Self-energy diagram on QCD side.}
    \end{center}

 The two-point Green's function reads:
\begin{eqnarray}
G(p) \;&=&\;{i\over \Sigma(p)} \nonumber\\\;&=&\;
  i\,{ A\,(\not\!p+m)
\,-\, 2B m \Delta \over 4m^2\Delta\left( A^2+A B- B^2 \Delta
\right)} \;. \label{G-2}
\end{eqnarray}

Now we first determine the effective lagrangian  up to order $1/m$
corrections using the matching condition (\ref{G-Gv}). With
(\ref{G-2}), the QCD side of (\ref{G-Gv}) reads:
\begin{eqnarray}
\mathrm{P}_{+} \;G(p)\; \mathrm{P}_{+} \;&=&\;i\, { A\,(2m+k\cdot
V) \;-\;2B m \Delta \over  4m^2\Delta\left( A^2+A B- B^2 \Delta
\right) }\;\mathrm{P}_{+}
 \nonumber\\ \bigskip
\;&=&\; {i\over c\,k\cdot V}\;\left(\, 1\,+\, {\mathbf{k}^2 \over
2mk\cdot V}  \,-\,{c_1+c_2+c_4 \over c }\;\delta\, \right) \;
\mathrm{P}_{+}\;, \label{G-2pp}
\end{eqnarray}
where $\delta\equiv k\cdot V /2m$ and $c=1+c_0+c_1$.

As we argued above, on the side of the effective theory,
contributions from the loop diagrams vanish. Thus all
contributions to the right side of the matching condition
(\ref{G-Gv}) arise from the tree diagrams. With all possible
insertion of higher order terms, the right side of the matching
condition (\ref{G-Gv}) reads:
\begin{eqnarray}
 {i\over Z\,k\cdot V}\;\left(\, 1\,+\, {\mathbf{k}^2 \over 2mk\cdot
V}  \,-\,{Z_e}\;\delta \,\right) \;\mathrm{P}_{+}\;.
\label{Gv-2pp}
\end{eqnarray}
Comparing (\ref{Gv-2pp}) to (\ref{G-2pp}), we see that
$Z=c=1+c_0+c_1$, and $Z_e=(c_1+c_2+c_4)/c$. With the one-loop
values given in (\ref{c0-4}), we have:
\begin{eqnarray}
Z\;&=&\; 1\,+\,{C_F\over 4\pi } \alpha_s(\mu)\Big(3 \ln{\mu^2\over
m^2}+4\Big )\,,
 \nonumber \\
Z_e\;&=&\;-{C_F\over 2 \pi} \alpha_s(\mu)\Big( 3\ln{\mu^2\over
m^2}+1\Big )\,. \label{z-ze}
\end{eqnarray}
These results are in agreement with those presented in
literatures. The coefficient $Z_m$ can only be determined by
matching the 3-point function in the next subsection.

We then use the matching condition (\ref{hmp}) to determine the
renormalized $h_{V-}(x)$ field up to order $1/m$ corrections. With
the two-point function given in (\ref{G-2}), the QCD side of
(\ref{hmp}) reads:
\begin{eqnarray}
\mathrm{P}_{-} \;G(p)\; \mathrm{P}_{+} \;&=&\;i\, {
A\,\mathrm{P}_{-} \not\!k\, \mathrm{P}_{+} \over 4m^2\,
\Delta\,\left( A^2+A B- B^2 \Delta \right) }
 \nonumber\\\bigskip \;&=&\;
 i\,{\mathrm{P}_{-}
\not\!k\, \mathrm{P}_{+}\over 2\,m\, c\,k\cdot V}\;\left[\; 1\,+\,
{\mathbf{k}^2 \over 2mk\cdot V} \,-\left(\,1+{c_2+c_4\over c
}-{c_1^2\over c(1+c_0) } \,\right)\,\delta \;\right]\;.
\label{G-2mp}
\end{eqnarray}

On the effective theory side, again contributions from loop
diagrams vanish while only tree diagrams survive.  Up to next
leading order correction terms five diagrams give nonzero
contributions, as shown in Fig.2.  The Feynman rules for the
operator insertions in these diagrams can easily be obtained from
(\ref{LV-1}) and (\ref{hm-1}). Their contributions to the right
side of (\ref{hmp}) reads:
\begin{eqnarray}
&&\;i\, {\mathrm{P}_{-} \not\!k\, \mathrm{P}_{+}\over 2m
}\;\left[\; {1\over Z\,k\cdot V}\,+\, {\mathbf{k}^2 \over
2mZ\,(k\cdot V)^2} \,- \,\left(\,{Z_e \over 2mZ
}\,+\,{d_1+d_2\over 2mZ } \,\right)\,\delta \;\right]\;.
\label{Gv-2mp}
\end{eqnarray}

   \begin{center}\begin{picture}(400,120)(0,0)
    \SetColor{Blue}
    \Line(0,73)(100,73)
    \Line(0,76)(100,76)
    \SetColor{PineGreen}
    \Line(46,71)(54,74.5)
    \Line(46,78)(54,74.5)
     \CCirc(100,74.5){4}{Magenta}{White}

    \SetColor{Blue}
    \Line(150,73)(250,73)
    \Line(150,76)(250,76)
    \SetColor{PineGreen}
    \Line(216,71)(224,74.5)
    \Line(216,78)(224,74.5)
   \CCirc(250,74.5){4}{Magenta}{White}
     \CBox(186,71)(193,78){Magenta}{Magenta}

    \SetColor{Blue}
    \Line(300,73)(400,73)
    \Line(300,76)(400,76)
    \SetColor{PineGreen}
    \Line(366,71)(374,74.5)
    \Line(366,78)(374,74.5)
   \CCirc(400,74.5){4}{Magenta}{White}
     \CBox(336,71)(343,78){Magenta}{White}

    \SetColor{Blue}
    \Line(50,8)(144,8)
    \Line(50,11)(143,11)
    \SetColor{PineGreen}
    \Line(96,6)(104,9.5)
    \Line(96,13)(104,9.5)
    \SetColor{Magenta}
    \Line(150,13)(142,13)
    \Line(142,13)(146,5)
    \Line(146,5)(150,13)

    \SetColor{Blue}
    \Line(250,8)(343,8)
    \Line(250,11)(344,11)
    \SetColor{PineGreen}
    \Line(296,6)(304,9.5)
    \Line(296,13)(304,9.5)
    \SetColor{Magenta}
    \Line(350,6)(342,6)
    \Line(342,6)(346,14)
    \Line(346,14)(350,6)
    \end{picture} \\ \bigskip\bigskip
    \begin{minipage}{5in} \footnotesize \baselineskip=14pt
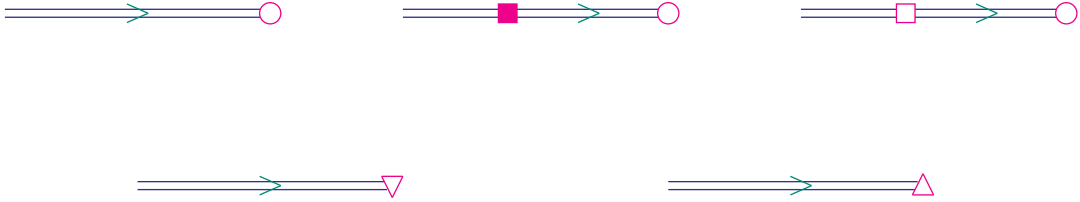
     {Fig. 2~ Diagrams contributing to the matching conditions on the
  effective theory side. The circle, up and down triangles represent
  the operators with coefficient $d_0$, $d_1$, and $d_2$,
  respectively, while the solid and the blank boxes represent the
  insertion of the kinetic and $(D\cdot V)^2/2m$ operators in the
  effective lagrangian.}
\end{minipage}
    \end{center}\bigskip\bigskip

Comparing (\ref{G-2mp}) to (\ref{Gv-2mp}), we obtain that:
\begin{eqnarray}
d_0\;&=&\;1\;, \nonumber \\
d_1+d_2 \;&=&\; 1\,-\,{c_1\over 1+c_0} \;. \label{d1d2}
\end{eqnarray}
We see that only the combination of the $d_1$ and $d_2$ can be
determined by the  matching through the 2-point function. To
determine them separately, we need to match the 3-point function.

\subsection{matching three-point function}
In this subsection, we determine the short-distance coefficients
$Z_m$ in (\ref{LV-1}) and $d_1$, $d_2$ in (\ref{hm-1}) by matching
the three-point function. Here the Feynamn diagrams with 3-gluon
vertex are involved. We use the background field
method\cite{background-field}, which can be used to simplify the
calculations significantly. In this method, QCD Ward identity for
the self-energy and the 1PI quark-gluon vertex takes a QED-like
form:
\begin{eqnarray}
k_\mu\,\Gamma^\mu(p_1,p_2) \;&=&\; \Sigma(p_2)\,-\,\Sigma(p_1)\;,
\label{ward}
\end{eqnarray}
where $\Gamma^\mu(p_1,p_2)$ is the 1PI 3-vertex with external
quark momenta $p_1$, and $p_2$, and gluon momentum $k=p_2-p_1$.

Up to next leading order correction terms, the general form of the
QCD 1PI vertex satisfying the Ward identity (\ref{ward}) with
quark near threshold can be written as:
\begin{eqnarray}
\Gamma^\mu(p_1,p_2) \;&=&\; \bar{A}\,\gamma^\mu \,+\,{\bar{B}
\over m } \, \bar{p}^\mu\,+\,{c_2\over 2m^2} \,\bar{p}^\mu
\,(\not\!\bar{p}-m) \;+\;{c_3 \over 4m }
\;[\,\not\!k,\gamma^\mu\,] \;, \label{gamma}
\end{eqnarray}
where
\begin{eqnarray}
\bar{A}\;&=&\;1+c_0 \,+\,c_2\,\bar{\Delta}\;, \nonumber \\
\bar{B}\;&=&\;c_1 \,+2\,c_4\,\bar{\Delta}\;,\nonumber \\
\bar{p}\;&\equiv&\; {1\over2} \,\left(\,p_1+p_2\,\right)\,,
\nonumber \\
\bar{\Delta} \;&\equiv&\; {p_1^2+p_2^2 -2m^2 \over 8m^2 } \;.
 \label{pab}
\end{eqnarray}

The one-loop coefficients $c_0$, $c_1$, $c_2$, and $c_4$ have been
given in (\ref{c0-4}). Thus we only need to evaluate the $c_3$.
For simplicity, we take the polarization vector of the gluon $e$
satisfying:
\begin{eqnarray}
&& e\cdot p_1 \;=\; e\cdot p_2 \;=\; e\cdot k \;=\; 0 \;.
\label{e}
\end{eqnarray}
Then we have:
\begin{eqnarray}
\Gamma^e(p_1,p_2) \;&\equiv&\; \Gamma^\mu(p_1,p_2)\, e_\mu \;=\;
\bar{A}\,\not\! e \;+\;{c_3 \over 4m } \;[\,\not\!k,\,\not\! e\,]
\;. \label{egamma}
\end{eqnarray}

In QCD, at one-loop level, there are two Feynman diagrams
contributing to $\Gamma^e(p_1,p_2)$ as shown in Fig. 3. A
straightforward calculation gives that:
\begin{eqnarray}
c_3(\mu)\;&=&\; \Big(\,2C_F+C_A\,\Big) \,{\alpha_s(\mu)\over
4\pi}\, \Big( \,\ln{\mu^2\over m^2}\,+\,2\,\Big)\;. \label{c3}
\end{eqnarray}

\bigskip
   \begin{center}\begin{picture}(400,80)(0,0)
    \SetWidth{0.7}
    \SetColor{YellowOrange}
    \Line(170,30)(30,30)
    \SetColor{Cerulean}
    \GlueArc(100,55)(50,210,330){5}{8}
    \SetColor{PineGreen}
    \Gluon(100,30)(100,70){5}{3}
    \SetColor{Brown}
     \Vertex(143.3,30){2} \Vertex(56.7,30){2} \Vertex(100,30){2}

    \SetColor{Blue}
    \SetWidth{0.7}
    \SetColor{YellowOrange}
    \ArrowLine(230,10)(370,10)
    \SetColor{Cerulean}
    \GlueArc(300,-15)(50,30,150){5}{8}
    \SetColor{PineGreen}
    \Gluon(300,40)(300,70){5}{2}
    \SetColor{Brown}
   \Vertex(343.3,10){2} \Vertex(256.7,10){2} \Vertex(300,40){2}
   \end{picture} \\\bigskip Fig. 3~
    Vertex diagrams on QCD side\end{center}

In the matching procedure, taking the polarization vector $e$ can
also significantly simplifies the calculations. In QCD, the
general form of the 3-point Green's function with this vertex
contributing to the matching conditions (\ref{G-Gv}) and
(\ref{hmp}) is expressed by:
\begin{eqnarray}
G^e(p_1,p_2) \;&=&\; G(p_1) \,\Gamma^e(p_1,p_2) \,G(p_2) \;.
\label{g12}
\end{eqnarray}

Now we first determine $Z_m$ by using (\ref{G-Gv}).  The QCD side
of (\ref{G-Gv}) reads:
\begin{eqnarray}
&&\; \mathrm{P}_{+}\,G(p_1) \,\Gamma^e(p_1,p_2) \,G(p_2)\,
\mathrm{P}_{+}\nonumber \\
 \;&=&\; -\, \mathrm{P}_{+}\,
 { A_1\,(\not\!k_1+2m) \;-\; 2B_1 m \Delta
   \over 4m^2\Delta_1\left( A_1^2+A_1 B_1- B_1^2 \Delta_1 \right)}
\,\left(\,\bar{A}\,\not\! e \;+\;{c_3 \over 4m }
\;[\,\not\!k,\,\not\! e\,] \,\right)\,
 \nonumber \\ \bigskip \;&&\;
 ~~~~\times\; { A_2\,(\not\!k_2+2m) \;-\; 2B_2 m \Delta_2
   \over 4m^2\Delta_2\left( A_2^2+A_2 B_2- B_2^2 \Delta_2 \right)}\,
\mathrm{P}_{+} \label{gpp-3}\;,
\end{eqnarray}
where the subscript 1 and 2 denote the momentum being $p_1$ and
$p_2$, respectively. Expanding it to leading order of $k$, we have
\begin{eqnarray}
-\,{1\over \,4\,m\,\,c^2\, k_1\cdot V\, k_2\cdot
V\,}\;\left(\,1+c_0+c_3\,\right)\;\mathrm{P}_{+}\,[\,\not\!k,\,\not\!
e\,] \, \mathrm{P}_{+}\;.\label{gpp-3k}
\end{eqnarray}
On the effective theory side, only the insertion of the color-
magnetic dipole term gives non-zero contribution for the gluon
polarization vector satisfying (\ref{e}). It reads:
\begin{eqnarray}
&&-\,{Z_m\over 4\,m\,Z\, k_1\cdot V\, k_2\cdot
V\,}\;\;\mathrm{P}_{+}\,[\,\not\!k,\,\not\! e\,] \,
\mathrm{P}_{+}\;.\label{gpp-3kv}
\end{eqnarray}
Comparing (\ref{gpp-3k}) to (\ref{gpp-3kv}), we determine that
\begin{eqnarray}
Z_m\;&=&\; {1+c_0+c_3 \over c}\;=\; 1 \,+\, {c_3-c_1 \over c}
\;.\label{gpp-zm}
\end{eqnarray}
With the coefficients given in (\ref{c0-4}) and (\ref{c3}), $Z_m$
at one-loop level reads:
\begin{eqnarray}
Z_m\;&=&\; 1\,+\,{\alpha_s(\mu)\over 4\pi} \left(\, C_A \,
\ln{\mu^2\over m^2}\,+\, 2C_A   \,+\, 2 C_F \,\right)
\;.\label{zm-1}
\end{eqnarray}

We then  determine $d_1$ and $d_2$ in the renormalized $h_{V-}$
field (\ref{hm-1}) by using the matching condition (\ref{hmp}).
The QCD side of (\ref{hmp}) reads:
\begin{eqnarray}
&&\; \mathrm{P}_{-}\,G(p_1) \,\Gamma^e(p_1,p_2) \,G(p_2)\,
\mathrm{P}_{+}\nonumber \\
\;&=&\;-\, \mathrm{P}_{-}\,
 { A_1\,(\not\!k_1+2m) \;-\; 2B_1 m \Delta
   \over 4m^2\Delta_1\left( A_1^2+A_1 B_1- B_1^2 \Delta_1 \right)}
\,\left(\,\bar{A}\,\not\! e \;+\;{c_3 \over 4m }
\;[\,\not\!k,\,\not\! e\,] \,\right)\, \nonumber \\
\;& & \;\times\; { A_2\,(\not\!k_2+2m) \;-\; 2B_2 m \Delta_2
   \over 4m^2\Delta_2\left( A_2^2+A_2 B_2- B_2^2 \Delta_2 \right)}\,
\mathrm{P}_{+} \label{gmp-3}\;,
\end{eqnarray}
The expression can be expanded as power series of $k$'s. Keeping
only the leading corrections, (\ref{gmp-3}) reads:
\begin{eqnarray}
\;&&\; \mathrm{P}_{-}\, \not\! e\, \mathrm{P}_{+}\;{1 \over
2m\,c\,k_2\cdot V} \,\left[\;1\,+{\mathbf{k}_2^2 \over
2m\,k_2\cdot V } \right.
\nonumber \\
\;&&\; -\,\left.
 \left(\,1\,-\,{2c_1-c_2-2c_3 \over 2(1+c_0)}\,\right)\,\delta_1
 \,-\,{c_1+c_2+c_4 \over c}\,\delta_2
 \,+\, {c_2+2c_3 \over 2(1+c_0)} \,\delta_2\;\right]
\nonumber \\
\;&&\;-\;
 \;{1+c_0+c_3 \over 4\,m\,c^2 }\,
 {1\over k_1\cdot V\, k_2 \cdot V }\;
 \mathrm{P}_{-}\not\!k\, \mathrm{P}_{+}
\;[\,\not\!k,\,\not\! e\,] \,\mathrm{P}_{+}\;.
 \label{gmp-4}
\end{eqnarray}
On the effective theory side, contributions may arise the
insertions of the operators both in $h_{V-}(x)$ given in
(\ref{hm-1}) and in the effective lagrangian given in
(\ref{LV-1}). With the polarization vector $e$, there are 6
Feynman diagrams contributing to it as shown in Fig.4. With
appropriate Feynman rules, they read:
\begin{eqnarray}
\;&&\; \mathrm{P}_{-}\, \not\! e\, \mathrm{P}_{+}\;{1 \over
2m\,Z\,k_2\cdot V} \,\left[\;1\,+{\mathbf{k}_2^2 \over
2m\,k_2\cdot V }
 -\, d_1\,\delta_1
 \,-\,{Z_e \over Z  }\,\delta_2
 \,-\, {d_2} \,\delta_2\;\right]
\nonumber \\
\;&&\;-
 \,{Z_m \over 4\,m\,Z }\,
 {1\over k_1\cdot V\, k_2 \cdot V }\;
 \mathrm{P}_{-}\not\!k\, \mathrm{P}_{+}
\;[\,\not\!k,\,\not\! e\,]\,\mathrm{P}_{+} \;.
 \label{gvmp-3}
\end{eqnarray}
\vspace{-20pt}
\begin{center}\begin{picture}(400,220)(0,0)
    \SetColor{Blue}
    \Line(0,123)(100,123)
    \Line(0,126)(100,126)
    \SetColor{PineGreen}
    \Line(46,121)(54,124.5)
    \Line(46,128)(54,124.5)
    \SetWidth{0.7}
    \Gluon(50,175)(100,126){5}{5}
     \SetWidth{0.5}
    \CCirc(100,124.5){4}{Magenta}{White}

    \SetColor{Blue}
    \Line(150,123)(250,123)
    \Line(150,126)(250,126)
    \SetColor{PineGreen}
    \Line(216,121)(224,124.5)
    \Line(216,128)(224,124.5)
    \SetWidth{0.7}
    \Gluon(200,175)(250,126){5}{5}
    \SetWidth{0.5}
   \CCirc(250,124.5){4}{Magenta}{White}
     \CBox(187,121)(194,128){Magenta}{Magenta}

    \SetColor{Blue}
    \Line(300,123)(400,123)
    \Line(300,126)(400,126)
    \SetColor{PineGreen}
    \Line(366,121)(374,124.5)
    \Line(366,128)(374,124.5)
    \SetWidth{0.7}
    \Gluon(350,175)(400,126){5}{5}
    \SetWidth{0.5}
    \CCirc(400,124.5){4}{Magenta}{White}
   \SetColor{Magenta}
     \CBox(337,121)(344,128){Magenta}{White}

    \SetColor{Blue}
    \Line(0,8)(94,8)
    \Line(0,11)(93,11)
    \SetColor{PineGreen}
    \Line(46,6)(54,9.5)
    \Line(46,13)(54,9.5)
    \SetWidth{0.7}
    \Gluon(50,60)(96,13){5}{5}
    \SetWidth{0.5}
    \SetColor{Magenta}
    \Line(100,13)(92,13)
    \Line(92,13)(96,5)
    \Line(96,5)(100,13)

    \SetColor{Blue}
    \Line(150,8)(243,8)
    \Line(150,11)(244,11)
    \SetColor{PineGreen}
    \Line(196,6)(204,9.5)
    \Line(196,13)(204,9.5)
    \SetWidth{0.7}
    \Gluon(200,60)(246,13){5}{5}
    \SetWidth{0.5}
     \SetColor{Magenta}
    \Line(250,6)(242,6)
    \Line(242,6)(246,14)
    \Line(246,14)(250,6)

    \SetColor{Blue}
    \Line(300,8)(400,8)
    \Line(300,11)(400,11)
    \SetColor{PineGreen}
    \Line(326,6)(334,9.5)
    \Line(326,13)(334,9.5)
    \SetWidth{0.7}
    \Gluon(325,60)(365,15){5}{5}
    \SetWidth{0.5}
    \CCirc(400,9.5){4}{Magenta}{White}
     \COval(365,9.5)(5,2)(0){Magenta}{Magenta}

    \end{picture} \\
    \bigskip
   \begin{minipage}{5in}\footnotesize \baselineskip=14pt
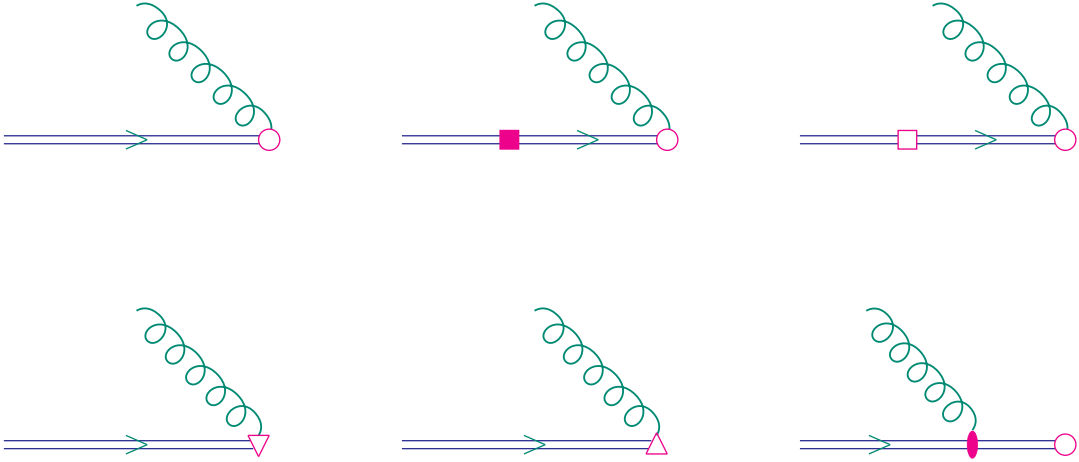
    Fig. 4~ Diagrams contributing to the matching conditions on the
 effective theory side. The notations are the same as figure 2. The
 solid Oval represents the insertion of the color-magnetic dipole
 operator.
  \end{minipage}
    \end{center}
\bigskip\bigskip
Comparing (\ref{gmp-3}) to (\ref{gvmp-3}), we determine that
\begin{eqnarray}
d_1 \;&=&\; 1\,-\,{2c_1-c_2-2c_3 \over 2(1+c_0)}\;,
\nonumber\\
d_2 \;&=&\; -\,{c_2+2c_3 \over 2(1+c_0)} \;. \label{d12}
\end{eqnarray}
These values are consistent with (\ref{d1d2}) obtained by matching
the 2-point function.

Substituting the short-distance coefficients in (\ref{c0-4}) and
(\ref{c3}) into (\ref{d12}),  we obtain the one-loop renormalized
coefficients for $d_1$ and $d_2$:
\begin{eqnarray}
 d_1(\mu) \;&=&\; \; 1\,+\, {\alpha_s(\mu) \over 4\pi } \;
  \Big[ \,\Big(\,C_A -2 C_F \,\big) \,\ln {\mu^2 \over m^2}
  \,-\,2C_F\,+\,2C_A \, \Big]\,,
 \label{d1}\\
 d_2(\mu) \;&=&\; -\,{\alpha_s(\mu)\over 4\pi} \,C_A \,
 \Big( \,\ln{\mu^2\over m^2}\,+\,2 \,\Big )\,,
\label{d2}
\end{eqnarray}
with  $C_A=3$. These are the central results of this section.
These results are in disagreement with that obtained by Balzereit
in \cite{Balzereit}. Since we use the same definition of the heavy
quark field, both result should be equal. The results presented in
this paper are derived rigorously using the matching conditions
while Balzereit obtained indirectly from the requirement of the
invariance of the effective lagrangian which is much more
complicated.

\section{RPI of the renormalized effective lagrangian}
In this section, I compare the renormalized transformation
determined by the matching condition (\ref{hmp}) with those given
in \cite{Kilian:1994mg}. I show that their results can easily be
understood by constructing the effective lagrangian in an
alternative way, in which an effective theory in four-component
field is constructed first, followed by its reduction to the
effective theory in two-component field. I then prove that  the
renormalized transformation determined by the matching conditions
(\ref{hmp}) can be written as the same form with the
transformation (\ref{D-V}) and (\ref{hm-tree}) with the covariant
derivative substituted by the operator which may be called as the
generalized covariant derivative. It means that the result
presented in this paper is consistent with that given in
\cite{Kilian:1994mg}.  Finally, I will show that the renormalized
effective lagrangian is reparameterization invariant under the
renormalized transformation.

\subsection{effective lagrangian in four-fermion field}
In the conventional method, a renormalized effective lagrangian is
constructed by following steps. First a proper field to describe
the low energy particles is chosen.  In HQET and NRQCD, this
effective field for describing the the heavy quark is just the
two-component field. Then the effective lagrangian in this field
is expanded as sum of local operators in terms of appropriate
counting rules. Then the renormalized short distance coefficients
of these local operators are determined by matching the full
theory and the effective theory. We refer this method as matching
after expansion.

Here we introduce an alternative way to determine the renormalized
effective lagrangian. In this method, renormalized local operators
expressed in the field of the full theory are added to the
lagrangian of the full theory by matching conditions. Then it is
expanded in terms of the two-component field. We refer this method
as matching before expansion.

Let's illustrate how this works in a hard cut-off regularization.

As  in the last section, we take different hard cut-off
regularization energy scales $\Lambda_e$ and $\Lambda_f$ in the
effective theory and in the full theory, respectively. They
satisfy $\Lambda_e \ll m \ll \Lambda_f$. In calculating the
one-loop 1PI diagrams in full QCD theory, we need to calculate the
loop momentum integrals from zero to $\Lambda_f$. They can be
separated into integrals from 0 to $\Lambda_e$ and integrals from
$\Lambda_e$ to $\Lambda_f$. The first part is just the same with
that in the effective theory while the second part gives extra
contributions. As argued above, the contributions from this region
can be written as local terms of external momenta and can be
expressed as contributions from local operators. Therefore, once
those local operators are added to the lagrangian, the effective
theory with hard cutoff $\Lambda_e$ can produce the same result of
the full theory with cutoff $\Lambda_f$. This argument can easily
be generalized to the case of multi-loops.

At this stage, those local operators are written in terms of Dirac
four-component field. A general form of the renormalized effective
lagrangian density with hard cutoff $\Lambda_e$ for heavy quark
field can formally be expressed as:
\begin{eqnarray}
{\cal L}_{\rm eff} \;&=&\; \bar{\Psi}(x) \, (i{\not D}-m) \,
\Psi(x) \,+\, \bar{\Psi}(x) \, O_1(x) \, \Psi(x)\;, \label{L-O1}
\end{eqnarray}
where $D^\mu=\partial^\mu - i g A^a_\mu T^a $ is the covariant
derivative. It  may  be denoted as
\begin{eqnarray}
{\cal L}_{\rm eff} \;&=&\; \bar{\Psi}(x) \, O(x) \, \Psi(x)\;,
\label{L-O}
\end{eqnarray}
for short by defining $O(x)\equiv i\not \! D-m + O_1(x)$.

The first term in (\ref{L-O1}) is just the tree-level lagrangian
while the second term arises from the renormalization with a
cut-off $\Lambda_e \ll m$. The operators in this term are
generally the function of the covariant derivative and the heavy
quark mass. It may contain terms such as $D^2+ m^2$, and
$g_sG^{\mu\nu} = i[D^\mu,D^\nu]$, which are suppressed by the
off-shell momentum of the heavy quark or the momenta of the
external gluons. They can be organized via appropriate power
counting rules. In perturbative calculations, the loop momentum
integral is from zero to $\Lambda_e$. In this region, both the
external and the loop momenta are smaller than $m$, hence the
$1/m$ expansion is allowed and the quark mass dependence is
extracted explicitly. Thus the energy scale $m$ is no longer
involved in the effective theory. I emphasis here that the
effective lagrangian density in this form is independent of the
velocity parameter $V$. Thus it automatically satisfies the RPI.

Higher dimensional operators appear in $O_1(x)$. It implies that
power divergences arise in the loop momentum integrals. In the
full theory the power divergences cancel when both the
contributions from quark and antiquark are included. However, in
the effective theory when we impose a hard cutoff $\Lambda_e\ll m$
on the loop momentum integrals, the contributions from antiquark
are excluded so that the power divergences do not cancel.
Nevertheless, those power divergences are artificial since the
power divergences from the diagram calculations just cancel those
from the short distance coefficients.

At one loop and leading order of $1/m$, the most general form of
the four-component effective lagrangian is
\begin{eqnarray}
{\cal L}_{\rm eff} \;&=&\; (1+c_0) \,\bar{\Psi}(x) \, (\,i{\not\!
D}-m\,) \,\Psi(x)
 \,-\, {c_1 \over 2m} \bar{\Psi}(x) \, (\,D^2+m^2\,) \, \Psi(x)\nonumber\\
 &&
 \;-\, {i c_2 \over 8m^2 } \,\bar{\Psi}(x)\,\Big[\,
 (\,i\not\!D-m\,)\,
 (\,D^2+m^2\,)\,+\,(\,D^2+m^2\,)\,(\,i\not\!D-m\,)\,\Big]\,\Psi(x) \nonumber\\
 &&
\;+\, {c_3 \over 4m} \,\bar{\Psi}(x) \, g_s\,\sigma^{\mu\nu}
G_{\mu\nu} \, \Psi(x) \,+\, {c_4\over 8m^3}\,\bar{\Psi}(x) \,
(\,D^2+m^2\,)^2 \, \Psi(x)
  \;. \label{L-1}
\end{eqnarray}
Calculating the the 1PI diagrams shown in Fig. 1 and Fig 3 using
this effective lagrangian and full QCD, we see that these
coefficients are just those $c_1-c_4$ given in (\ref{c0-4}), and
(\ref{c3}).

\subsection{effective lagrangian in two-component field}

Now let's reduce (\ref{L-1}) to the effective lagrangian in
two-component field. The equation of motion now reads:
\begin{eqnarray}
\mathrm{P}_{-} \;\overline{O}(x) \;
       \left( \,h_{V+}(x) \,+ \,h_{V-}(x)\,\right) \;&=&\; 0\;,
  \label{hpm}
\end{eqnarray}
where $\overline{O}(x)$ is the $O(x)$ whose covariant derivative
$iD$ is replaced  by $iD+mV$ due to the phase factor in the field
redefinition. It can be regarded as the renormalized equation of
motion.

From Eq.(\ref{hpm}), we can express $h_{V-}(x)$ as a function of
$h_{V+}(x)$ formally as:
\begin{eqnarray}
 h_{V-}(x) \;&=&\; \frac{1}{2m+iD(x)\cdot V
 \,-\,\mathrm{P}_{-}\overline{O}_1(x)\mathrm{P}_{-}}\,
  \mathrm{P}_{-} \,\left(\,i\not{D} \,+\,\overline{O}_1(x)\,\right)
  \, h_{V+}(x)\;,
 \label{hm}
\end{eqnarray}
where $\overline{O}_1(x)$ is the $O_1(x)$ whose covariant
derivative $iD$ is substituted by $iD+mV$. This modifies the tree
level expression (\ref{hm-tree}).
 Once the form of $O_1(x)$ is
given, the right hand side of Eq.(\ref{hm}) can be expanded as
power series of $1/m$. With $O_1(x)$ given in (\ref{L-1}), up to
order $\alpha_s$ and $1/m^2$, $h_{V-}(x)$ reads:
\begin{eqnarray}
 h_{V-}(x) \;&=&\; \mathrm{P}_{-} \,\Big[ \;{1 \over 2m}\,i\not \!D \,+\,
 {1\over 4m^2} \, \left(\,1\,-\, {2c_1-c_2 -2c_3 \over 2(1+c_0)
 }\,\right)\,
   D\cdot V  \not \!D  \nonumber \\
 &&   \,-\, { c_2  + 2c_3 \over 8(1+c_0)m^2}
   \,\not \!D  D\cdot V \;\Big]\,
    h_{V+}(x)\;.
 \label{hm-2}
\end{eqnarray}
Comparing this with (\ref{hm-1}), (\ref{d12}), we see that they
are in agreement.

Finally, with equation of motion (\ref{hpm}), the effective
lagrangian (\ref{L-O}) is reduced to:
\begin{eqnarray}
{\cal L}_{\rm eff} \;&=&\;
 \bar h_{V+}(x) \,
  \overline{O}(x)\,\Big(\, h_{V+}(x) \,+\, h_{V-}(x)\,\Big)\nonumber \\
  \;&=&\;
 \Big(\,\bar h_{V+}(x) \,+\,\bar h_{V-}(x)\,\Big)\,
  \overline{O}(x)\,h_{V+}(x) \;.
  \label{L-O-h}
\end{eqnarray}
This is just the two-component effective lagrangian. It can be
expanded as power series of $1/m$. In this way, the four component
effective lagrangian is reduced to the two component effective
lagrangian. Up to order $\alpha_s$ and $1/m$ correction, it
reduced to the effective lagrangian (\ref{LV-1}). Therefore, the
effective lagrangian obtained by these two different approaches
are just the same.

\subsection{comparison with the previous work}
In Sec. II, the matching condition for determining the
renormalized $h'_{V-}(x)$ field is given by Eq.~(\ref{hmp}). In
the last subsection,  the $h_{V-}(x)$ field has been obtained by
equation of motion. Its expression is given by Eq.~(\ref{hm}). In
this subsection, I will show that the  small component fields
obtained by these two different methods are identical. They
uniquely determine the renormalized transformation of the heavy
quark field against the infinitesimal variation of the velocity
parameter $V$. Adding both sides of (\ref{hpp}) and (\ref{hmp})
together, we have
\begin{eqnarray}
 \langle \, 0\, |\,T\,\Big( \,h_{V+}(x)\,+\,h'_{V-}(x)\,\Big)
 \,\bar{h}_{V+}(y) \, |\, 0 \, \rangle^B
 \; &\doteq & \;
 \, \langle \,0\, |\, T\,\Psi(x)\,\bar{\Psi}(y)\,|\, 0\, \rangle^B
 \,\mathrm{P}_{+}
 \label{hpmp} \,.
\end{eqnarray}
where $\langle 0 | T\Psi(x)\,\bar{\Psi}(y)| 0 \rangle^B$ is a full
propagator under arbitrary external field $B^\mu(x)$. It is
satisfied order by order in $\alpha_s$. Suppose we calculate the
left hand side at tree level with the renormalized effective
lagrangian. To validate this equation, the right hand side then
should also be calculated to the tree level with containing
contributions of the loop momentum integrals from $\Lambda_e$ to
$\Lambda_f$. This can be calculated by the renormalized
four-component effective lagrangian (\ref{L-O1}) to tree level.
Thus it satisfies the following equation
\begin{eqnarray}
  O(x)\,G(x,y;B)\;&=&\;i\,\delta^4 (x-y)
 \label{O-G}  \;.
\end{eqnarray}
 Acting an operator $\mathrm{P}_{-}\overline{O}(x)$ on the left hand side
and $\mathrm{P}_{-}{O}(x)$ on the right hand side of (\ref{hpmp}),
the right hand side vanishes immediately since
$\mathrm{P}_{-}\cdot \mathrm{P}_{+}=0$. Since we only calculate
them at tree level, the operator $\overline{O}(x)$ can be moved
within the bracket:
\begin{eqnarray}
 \mathrm{P}_{-}\,\langle \, 0 \,|T\,\big[\,O(x)\,\big(\, h_{V+}(x)
 \,+\,h'_{V-}(x)\,\big)\,
  \bar{h}_{V+}(y)\,\big]\,|\, 0\, \rangle^B
  \;& = &\; 0
 \label{O-hpm} \;.
\end{eqnarray}
Since the argument $y$ in $\bar{h}_{V+}(y)$ is arbitrary and this
correlation function contains interaction with arbitrary
background gluon field, the unique solution of this equation is
\begin{eqnarray}
 \mathrm{P}_{-} \,\overline{O}(x)\,\Big(\, h_{V+}(x)\,+\,h'_{V-}(x)
 \,\Big) \;& = &\; 0
 \label{O-hpm1} \,.
\end{eqnarray}
This is just identical to (\ref{hpm}) if $h'_{V-}(x)$ is the same
with $h_{V-}(x)$. This implies that the renormalized $h'_{V-}(x)$
determined by the matching condition (\ref{hmp}) is identical to
that from the equation of motion (\ref{hpm}).

\subsection{RPI of the renormalized effective lagrangian}

In this subsection, let's prove that the renormalized effective
lagrangian (\ref{L-O-h}) is invariant under the transformation
(\ref{D-V}) or (\ref{D-h-pm}) with renormalized small component
field.

It follows that from an infinitesimal transformation of the
effective lagrangian (\ref{L-O-h})
\begin{eqnarray}
{\Delta \cal L}_{\rm eff} \; & \doteq & \;
 \Delta \bar h_{V+}(x) \,
  \overline{O}(x) \,h_{V}(x) \,
 + \,\bar h_{V+}(x) \,
  \overline{O}(x)\, \Delta h_{V}(x) \nonumber\\
 \;& = &\;
  \bar h_{V}(x) \, \frac{\DVslash}{2}\,
  \overline{O}(x) \,h_{V}(x)
 \,+ \, \bar h_{V+}(x) \,
 \overline{O}(x) \, \Delta\,  h_{V}(x)
  \;.
  \label{D-L-1}
\end{eqnarray}
We have use a shorthand notation $h_V(x)= h_{V+}(x) + h_{V-}(x)$.
It is emphasized here that the operator $O(x)$ which is from the
four component effective field theory is invariant against the
variation of the velocity $V$. Any change arising from the phase
factor in the definition of the effective field has been omitted
simply because it is trivial under the transformation.

Imposing an infinitesimal transformation on the equation of motion
(\ref{hpm}), we obtain that
\begin{eqnarray}
 - \frac{\DVslash}{2} \, \overline{O}(x) \, h_{V}(x) \,+\, \mathrm{P}_{-}
 \,\overline{O}(x)\, \Delta h_{V}(x)
 \; & \doteq&\; 0 \;.
  \label{D-hpm}
\end{eqnarray}
With it, (\ref{D-L-1}) can be rewritten as
\begin{eqnarray}
{\Delta \cal L}_{\rm eff} \;& \doteq &\;
  \bar h_{V}(x)\,\overline{O}(x) \, \Delta h_{V}(x)
   \;.
  \label{D-L-3}
\end{eqnarray}
Notice that $\mathrm{P}_{+}\, h_{V-}(x)=0$. Imposing an
infinitesimal on it, we immediately have
\begin{eqnarray}
\mathrm{P}_{+} \,\Delta h_{V-}(x) \;&=&\; -\,\frac{\DVslash}{2}\,
h_{V-}(x) \;.
\end{eqnarray}
Adding it together with $\mathrm{P}_{+}\Delta h_{V+}(x) = \DVslash
/2 \; h_{V-}(x) $, we have
\begin{eqnarray}
\mathrm{P}_{+}\,\Delta h_{V}(x)\; = \;0. \;,
\end{eqnarray}
With it, (\ref{D-L-1}) is  reduced to
\begin{eqnarray}
{\Delta \cal L}_{\rm eff} \;& \doteq &\;
  \bar h_{V}(x)\,\overline{O}(x) \,\mathrm{P}_{-}\, \Delta h_{V}(x)
   \;.
  \label{D-L-4}
\end{eqnarray}
It follows that ${\Delta \cal L}_{\rm eff} =0$ from the equation
of motion $ \bar h_{V}(x)\,O(x) \,\mathrm{P}_{-} =0 $. Thus we
have shown that the renormalized effective lagrangian
(\ref{L-O-h}) is invariant against the variation of the velocity
parameter $V$ under the infinitesimal transformation (\ref{D-V})
with the renormalized small component field.

\section{conclusion}
The RPI is an important theoretical issue in the heavy quark
effective theory and the NRQCD effective theory. It is required by
the consistence of the effective theory. It also leads to
interesting
applications\cite{Neubert:1993iv}\cite{Chen:1993us}\cite{Chen:1994dg}.
The transformation of heavy quark field under the variation of the
velocity parameter $V$ proposed by Chen\cite{Chen:sx} with tree
level expression of the small-component field keeps the tree level
effective theory invariant. However, at loop level, the
transformation needs to be renormalized in the renormalized
effective theory. In this paper, I show that the renormalized
transformation of the heavy quark keeps the same form as Chen's
transform while the small component field needs to be
renormalized. I derive the matching conditions for determining the
renormalized transformation by imposing an infinitesimal
transformation on the relations between the Green's functions in
the full QCD and those in the effective theory. These matching
conditions are essential for studying the renormalization issue in
RPI.  As an application of these matching conditions, I determine
the renormalized transformation up to order $1/m^2$. I also show
that the previous result in \cite{Kilian:1994mg} can be understood
clearly by building the effective theory in an alternative way, in
which the a renormalized effective lagrangian in Dirac
four-component field is constructed first, followed by its
reduction  to the two-component effective lagrangian. The
renormalized small component field is then obtained by the
equation of motion. The four-component effective lagrangian
automatically satisfies RPI. Thus RPI  cannot give any constraints
on any operators in it. When it is reduced to the two-component
effective theory, the same operator with certain coefficient may
appear in different terms. The RPI can be used to connect those
terms. I also show that the obtained renormalized small component
fields by these two methods turn out to be equivalent while the
matching conditions provide a systematic way to determine the
renormalized transformation  to any desired order in $1/m$ and
$\alpha_s$ expansions..


\appendix
\section{generating functional of Green's functions} In this
appendix, we derive the relations between the Green's functions in
QCD full theory and that in the effective theory using generating
functional method. It is similar with that given in
\cite{Mannel:1991mc} and \cite{Chen:sx}. We use the background
field method\cite{background-field,abbott} for gluon field
interactions, which preserves explicitly the gauge covariant.

In QCD full theory, the generating functional reads
\begin{eqnarray}
 Z[\eta,\bar{\eta},J,B] &=&
 \int \delta[\eta,\bar{\eta},A]\exp i\int d^4x ( I_Q(x) +I_g(x))\;,
 \label{Z-qcd}
\end{eqnarray}
where  $\eta$, $\bar{\eta}$,$J$ are the external sources for heavy
quark, antiquark and gluon field, $B$ is the background gluon
field,  $I_g$ is given by
\begin{eqnarray}
I_g(x) &=& - {1\over 4} F^a_{\mu\nu}F^{a\mu\nu}
      - {1\over 2\xi}(G^a)^2
      + \ln \det\Big[{\delta G^a \over \delta \omega^b}\Big]
      +J_\mu^a A^{a\mu }  \;,
\end{eqnarray}
with
\begin{eqnarray}
F^a_{\mu\nu} &=& \partial_\mu (A + B)_\nu^a
                -\partial_\nu (A + B)_\mu^a
                +gf^{abc}(A + B)_\mu^b (A + B)_\nu^c \;,
\end{eqnarray}
\begin{eqnarray}
 G^a &=& \partial_\mu A_\mu^a
                +gf^{abc} B_\mu^b A^{c\mu}
                \;,
\end{eqnarray}
being the gauge-fixing term. If $J_\mu$ satisfies the following
relation
\begin{eqnarray}
 {\delta W \over \delta B^a_\mu}
 + \int d^4y \Big[
 {\delta {W} \over \delta J^b_\nu}
 {\delta J^b_\nu(y) \over\delta B^a_\mu} \Big]
 &=& -J_\mu^a\;,
\end{eqnarray}
with $W[\eta,\bar{\eta},J,B] = - i \ln Z[\eta,\bar{\eta},J,B]$,
$W[\eta,\bar{\eta},J,B]$ is just the effective action regarding to
the gluon field $B$ with gauge-fixing term
\begin{eqnarray}
  G^a &=& \partial_\nu (A-B)_\nu^a
  +gf^{abc} B_\mu^b A_\nu^c   \;,
\end{eqnarray}
and $I_Q$ reads
\begin{eqnarray}
I_Q(x) &=& \overline{\Psi}(x) (i\not \!{D} -m) \Psi(x)
+\overline{\eta}(x) \Psi(x) +\overline{\Psi}(x) \eta (x) \;,
\end{eqnarray}
The quark field can be integrated out formally and then we have
\begin{eqnarray}
 Z[\eta,\bar{\eta},J,B]\,&=&\,
 \int \delta[A] \det[i\not\!D-m] \exp i\int d^4x ( I'_Q +I_g)\;,
 \label{Z-A}
\end{eqnarray}
where $I'_Q$ remains the same form as $I_Q$. But the quark field
now is related to the external source $\eta(x)$ by the following
equation of motions:
\begin{eqnarray}
(i\not\!D -m )\Psi(x) = -\eta(x)\;.
\end{eqnarray}
The generating functional of the effective theory is similar with
that of the full theory except the heavy quark action. The
effective lagrangian is substituted by (\ref{L-O}). In the
external source term of the heavy quark  only the large component
effective field defined in (\ref{hvp}) couples to the external
source. The action of the heavy quark is given by:
\begin{eqnarray}
 I^{V+}_Q(x) &=& \overline{\Psi}(x) O(x) \Psi(x)
 +\bar{\eta}(x)\mathrm{P}_{+} h_{V+}(x)
 +\bar{h}_{V+}(x)\mathrm{P}_{+}\eta (x) \;.
\end{eqnarray}
Similarly, integrating out the heavy quark field, the generating
functional takes the same form as (\ref{Z-A}) with the effective
action of the heavy section is substituted by %
\begin{eqnarray}
 I'^{V+}_Q(x) &=&
 \overline{h}_{V}(x) \overline{O}(x) h_{V}(x)
 +\bar{\eta}(x)\mathrm{P}_{+} h_{V+}(x)
 +\bar{h}_{V+}(x)\mathrm{P}_{+}\eta (x) \;,
 \label{I'V}
\end{eqnarray}
with $h_{V}(x)= h_{V+}(x) + h_{V-}(x)$.

 The quark field now is
related to the external source $\eta(x)$ by the following equation
of motions:
\begin{eqnarray}
 \overline{O}(x) h_{V+}(x)
 &=& - \mathrm{P}_{+}\eta (x) \;.
\end{eqnarray}
Multiplying $\mathrm{P}_{-}$ on both sides, the right hand
vanishes and we obtain the renormalized equation of motion:
\begin{eqnarray}
\mathrm{P}_{-} \;\overline{O}(x) \;
       \Big( h_{V+}(x) + h_{V-}(x)\Big) &=& 0\;.
\label{E-Q}
\end{eqnarray}
This is just the equation (\ref{hpm}).  The renormalized
$h_{V-}(x)$ can be related to $h_{V+}(x)$ by (\ref{hm}).  With the
equation of motion (\ref{E-Q}), (\ref{I'V}) can be simplified as
\begin{eqnarray}
I'^V_Q &=&
 \bar h_{V+} (x)\overline{O}(x) h_{V}(x)\,
 + \bar{\eta}(x)\mathrm{P}_{+} h_{V+}(x)
 +\bar{h}_{V+}(x)\mathrm{P}_{+}\eta (x) \;.
  \label{I'-V-1}
\end{eqnarray}
This gives the effective lagrangian density (\ref{L-O-h}).

 The quark determinant in (\ref{Z-A}) is responsible for the
contributions of the heavy quark loop. It is the same in the full
theory and in the effective theory and is suppressed by $1/m^2$ at
least. Thus we may ignore it.

The full quark propagator with background field $B^\mu(x)$ is
gained by differentiating over the external sources.
\begin{eqnarray}
 G(x,y;B) &=&
 {\delta^2 \over \delta\eta(x) \bar\delta \eta(y) }
 W(\eta,\bar\eta,J,B)\;.
\end{eqnarray}
If the hard cutoff energy scale is set to $\Lambda_f$, the same
with that in the QCD full theory, the $O(x)$ is then to be
$i\not\!D-m$, just the same with that in the full theory. The
effective lagrangian is just the nonlocal form (\ref{L-tree}). In
this case the only difference of  the effective theory and the
full theory is the external source term.  One immediately gains
the relations between the Green's functions of the full theory and
those in the effective theory (\ref{G-Gv}). This relation ensures
that the nonlocal effective theory is equivalent to the QCD full
theory. The local effective theory with the hard cutoff
regularization scale $\Lambda_e$ is equivalent to that the
nonlocal effective theory with hard cutoff $\Lambda_f$. This
ensures the validation of the relation (\ref{G-Gv}).



\begin{thebibliography}{widest-label}

\bibitem{HQET}
M.~A.~Shifman and M.~B.~Voloshin,
Sov.\ J.\ Nucl.\ Phys.\  {\bf 45}, 292 (1987) [Yad.\ Fiz.\  {\bf
45}, 463 (1987)];
M.~A.~Shifman and M.~B.~Voloshin,
Sov.\ J.\ Nucl.\ Phys.\  {\bf 47}, 511 (1988) [Yad.\ Fiz.\  {\bf
47}, 801 (1988)];
N.~Isgur and M.~B.~Wise,
Phys.\ Lett.\ B {\bf 232}, 113 (1989);
Phys.\ Lett.\ B {\bf 237}, 527 (1990);
H.~D.~Politzer and M.~B.~Wise,
Phys.\ Lett.\  {\bf 206B}, 681 (1988);
Phys.\ Lett.\  {\bf 208B}, 504 (1988).
E.~Eichten and B.~Hill,
Phys.\ Lett.\ B {\bf 234}, 511 (1990);
Phys.\ Lett.\ B {\bf 240}, 447 (1990);
B.~Grinstein,
Nucl.\ Phys.\ B {\bf 339}, 253 (1990);
A.~F.~Falk, H.~Georgi, B.~Grinstein and M.~B.~Wise,
Nucl.\ Phys.\ B {\bf 343}, 1 (1990);
H.~Georgi and M.~B.~Wise,
Phys.\ Lett.\ B {\bf 243}, 279 (1990);
M.~Neubert,
Phys.\ Rept.\  {\bf 245}, 259 (1994).

\bibitem{NRQCD}
W.~E.~Caswell and G.~P.~Lepage,
Phys.\ Lett.\ B {\bf 167}, 437 (1986).
G.~P.~Lepage and B.~A.~Thacker, Nucl.\ Phys.\ Proc.\ Suppl.\ {\bf
B4}, 199 (1988);

G.~T.~Bodwin, E.~Braaten and G.~P.~Lepage,
Phys.\ Rev.\ D {\bf 51}, 1125 (1995) [Erratum-ibid.\ D {\bf 55},
1125 (1995)].


\bibitem{Luke:1992cs}
M.~E.~Luke and A.~V.~Manohar,
Phys.\ Lett.\ B {\bf 286}, 348 (1992).

\bibitem{Manohar:1997qy}
A.~V.~Manohar,
Phys.\ Rev.\ D {\bf 56}, 230 (1997) [arXiv:hep-ph/9701294].

\bibitem{Chen:sx}
Y.~Q.~Chen,
Phys.\ Lett.\ B {\bf 317}, 421 (1993).


\bibitem{Finkemeier:1997re}
M.~Finkemeier, H.~Georgi and M.~McIrvin,
Phys.\ Rev.\ D {\bf 55}, 6933 (1997).

\bibitem{Kilian:1994mg}
W.~Kilian and T.~Ohl,
Phys.\ Rev.\ D {\bf 50}, 4649 (1994) [arXiv:hep-ph/9404305].


\bibitem{Sundrum:1997ut}
R.~Sundrum,
Phys.\ Rev.\ D {\bf 57}, 331 (1998).

\bibitem{Balzereit}
C.~Balzereit,
hep-ph/9809226 (unpublished).


\bibitem{Neubert:1993iv}
M.~Neubert,
Phys.\ Lett.\ B {\bf 306}, 357 (1993).

\bibitem{Chen:1993us}
Y.~Q.~Chen and Y.~P.~Kuang,
Z.\ Phys.\ C {\bf 67}, 627 (1995).


\bibitem{Chen:1994dg}
Y.~Q.~Chen, Y.~P.~Kuang and R.~J.~Oakes,
Phys.\ Rev.\ D {\bf 52}, 264 (1995).

\bibitem{Mannel:1991mc}
T.~Mannel, W.~Roberts and Z.~Ryzak,
Nucl.\ Phys.\ B {\bf 368}, 204 (1992).

\bibitem{background-field}
B.~S.~Dewitt,
Phys.\ Rev.\ {\bf 162}, 1195 (1967).

\bibitem{abbott}
L.~F.~Abbott,
Nucl.\ Phys.\ B {\bf 185}, 189 (1981).

\end{thebibliography}
\end{document}